**Casimir force between plane-layered structures:**
**van Kampen-Schram approach for finite temperature**

M.V. Davidovich
*Saratov State University, 410012 Saratov, Russia*
E-mail: davidovichmv@yandex.ru

Based on the van Kampen-Schram method, a formula for the Casimir force between two plane-layered structures is obtained, which includes the sum of the Matsubara frequencies and a contour integral. It is shown that the contour integral is not taken into account in the Lifshitz formula. Taking it into account gives a small contribution at high temperatures when the Lifshitz formula is approximately correct. However, at low temperatures, its contribution is significant, and the Lifshitz formula for the finite temperature does not transform into the Lifshitz formula at zero temperature. An example of interaction of metal plates is considered.

In [1], Casimir performed the summation of zero vacuum fluctuations $\hbar\omega_n/2$ in the cavity region $|z|<a/2$ of Fig. 1 (insert) with ideally conducting walls at zero temperature $T$=0. The total free energy $E=\sum_n \hbar\omega_n/2$ of the resonator turned out to be infinite, since the real frequencies $\omega_n = c\sqrt{k_{xk}^2+k_{yl}^2+k_{zm}^2}$ are unlimited and degenerate for E-modes and H-modes (with matching indices). Here $k_{xk}=k\pi/L$, $k_{yl}=l\pi/L$, $k_{zm}=m\pi/d$, and we denote the multiindex $n=(k,l,m)$, $k,l,m$ are integers that are not equal to zero at the same time. For magnetic modes (TE$_{klm}$) $m\geq 1$, and the case is $k=l=0$ is excluded. For electric modes (TH$_{klm}$) $k\geq 1$, $l>1$, $m\geq 0$. Next, Casimir moved from the double sum over the transverse indices to the integral by replacing $\sum_{k,l}(\ )\to(L^2/\pi^2)\int(\ )dk_x dk_y$ $dk=(L/\pi)dk_x$, $dl=(L/\pi)dk_y$ with $L\to\infty$. The resonator has become an open Fabry-Perrault resonator (RFP). Since the energy of zero fluctuations in it is also infinite, the infinite energy of zero fluctuations in the region $|z|>D$ at $D\to\infty$ was subtracted from it (see insert, Fig. 1). It corresponds to the energy of free space. In fact, this means subtracting the energy of the RFP at $a\to\infty$ which Casimir tacitly postulated. The difference also turned out to be infinite, which is due to the ideal conductivity. Since there are no media with ideal conductivity, and at low frequencies there is a connection between the surface conductivity of a metal $\sigma(\omega)$ and its dielectric constant (DP) according to the Drude formula) $\varepsilon(\omega)\approx -i\omega_p^2/(\omega\omega_c)=-i\sigma(0)\varepsilon_0/\omega$, infinite conductivity means the tendency of the plasma frequency (PF) $\omega_p$ to infinity, and at low frequencies the tendency to zero collision frequency (CP) $\omega_c \to 0$. We use the dependence $\exp(i\omega t)$ as usual in radio engineering and electrodynamics (to switch to the dependence $\exp(-i\omega t)$ often used in theoretical physics, the sign of the imaginary unit should be changed). To find the force between thick metal plates, Casimir assumed that at low frequencies they completely reflect, and in the transition to high frequencies they become completely transparent. He applied the Euler-Maclaurin formula, resulting in a finite energy difference $\Delta E/L^2 = -\hbar c\pi^2/(720a^3)$ and a specific force per unit area $F = \hbar c\pi^2/(240a^3)$. The attractive force means the external pressure of the field on the plates.

In [2] Lifshitz gave two formulas for the Casimir force:

$$F(d)=\frac{\hbar}{2\pi^2c^3}\int_0^\infty \xi^3 d\xi \int_1^\infty \sum_{\mu=e,h}\frac{r_\mu^2\exp(-2p\xi a/c)}{1-r_\mu^2\exp(-2p\xi a/c)}p^2dp\,, \tag{1}$$

$$F(d,T)=\frac{k_BT}{\pi}\sum_{n=0}^{\infty}\frac{k_n^3}{1+\delta_{n0}}\int_1^\infty \sum_{\mu=e,h}\frac{p^2dp}{r_{\mu n}^{-2}\exp(2pk_na)-1}\,. \tag{2}$$

The first formula refers to the case of zero temperature $T$=0. In it $\xi = i\omega$, and $r_\mu^2$ are the squares of the reflection coefficients from the plates on the gap side for E-waves ($p$- polarization, $\mu=e$), as well as for H-waves (s- polarization, $\mu=h$): $r_e=(s-\varepsilon p)/(s+\varepsilon p)$, $r_h=(s-p)/(s+p)$, $s=\sqrt{p^2-1+\varepsilon}$. In formula (2) $k_n = 2\pi k_B T n/(\hbar c) = 2\pi n/\delta_T$, $r_{\mu n} = r_\mu(k_n)$, where $\delta_T = \hbar c/(k_B T)$ is the characteristic thermal length. From formula (1) at $r_\mu = 1$ (ideal conductivity) follows the result of Casimir [2-4]). Formula (1) also describes forces in structures with dissipation. In an ideal RFP, there are no surface waves along the screens, and all modes are radiated. Formula (1) is also applicable to lossy structures in which there are damped surface waves and outgoing (radiated) volumetric modes.

In [5,6] (see also [3]), formula (1) was obtained using the “principle of argument” theorem by calculating the sum $E=\sum_n \hbar\omega_n/2$ where the frequencies were already the complex roots $\omega_n' \pm i\omega_n''$ of the characteristic equations $f_\mu(\omega,\kappa)=0$ of an arbitrary plane-layered structure, Fig. 1 (inset), symmetrically arranged relative to the imaginary axis. The method [5,6] is a generalization of formula (1). In particular, for the Lifshitz problem we have $f_\mu(k,\kappa) = r_\mu^{-2}(k,\kappa)\exp\left(2\tilde{k}_z d\right) - 1$. In the case $L \to \infty$, the characteristic equations for frequencies turn into dispersion equations (DE) for plasmon-polaritons (PP) along the surfaces of the structure (for their role in the dispersion forces, see [7,8]). Fast leakage (radiated) PP contributes mainly to the strength in the far zone, whereas slow (surface) PP attenuates and contributes to the force in the near zone. The DEs have complex roots in the upper half-plane of the complex frequency plane. The principle of the argument implements Wick rotation (see [3-6]) with a transition to a contour integral along the imaginary frequency axis and an infinite right semicircle [6]. The integral over it vanishes. A contour integral remains along the imaginary axis. The van Kampen-Schram method reduces to summing complex frequencies with a multiplier $2\pi i$, and the sum is expressed in terms of the sum of the semi-residuals and the remaining contour integral in the domain $(-i\infty, i\infty)$ [6], which can be reduced to integration over the domain $0 \le \xi < \infty$ in (1), $\xi = i\omega$. In this case, when rotating, the complex conjugate frequencies pass into the right half-plane and have opposite imaginary parts $\mp\omega_n''$ there, which they are compensated, and integral (1) becomes valid. Generalization (1) can be written as

$$F(d) = \frac{\hbar}{2\pi^2 c^3}\int_0^\infty \xi^3 d\xi \int_1^\infty \sum_{\mu=e,h} \frac{p^2}{f_\mu(p,\xi,d)} dp \,. \tag{3}$$

There is a discussion in the literature about the possibility of Casimir summation at complex frequencies [3,7,8]. It was supposed to take the real part of this amount. The absence of the need to take the real part of the energy at complex frequencies can be explained as follows. When structures are excited at a given frequency during dissipation in the monochromatic mode, energy per a period is lost (converted into heat), dissipating over the entire spectrum. However, this energy is replenished by the source of excitation, so the free energy of the field-matter system as a whole is conserved. In thermodynamic equilibrium, at each frequency, the body absorbs as much as it emits (due to thermal fluctuations). The total free energy of the field-matter system is conserved, so there are no losses, and it is not necessary to take the real part.

In [9], the Casimir method was extended to calculate the sum for the average energies of quantum oscillators, taking into account zero and thermal fluctuations in RFCs with ideally conducting walls: $E = \sum_n \hbar\omega_n \coth(x_n)/2$. Here $x_n = \hbar\omega_n/(2k_B T) = \omega_n\delta_T/2$. The Euler-Maclaurin formula was used to isolate the Casimir energy. Differentiating the energy by distance and using a similar approach [1], Mehra obtained the force at a low temperature by decomposing it into a small parameter ($a/\delta_T \ll 1$) in the form

$$F(a,\delta_T)=\frac{\hbar c\pi^2}{240a^4}\left[1-\frac{240}{\pi}\frac{a}{\delta_T}\exp(-\pi\delta_T/d)+\frac{16}{3}\left(\frac{a}{\delta_T}\right)^4\right], \tag{4}$$

and at high temperature ($a/\delta_T >> 1$) by summing by Poisson in the form

$$F(a,\delta_T)=\frac{\hbar c}{4\pi a^3\delta_T}\left[\varsigma(3)+\left(2+8\pi\frac{a}{\delta_T}+\left(4\pi\frac{a}{\delta_T}\right)^2\right)\exp\left(-\frac{4\pi a}{\delta_T}\right)\right]. \tag{5}$$

In this case (5) approximately is

$$F(a,\delta_T)=\frac{\hbar c}{4\pi a^3\delta_T}\left(\varsigma(3)+16\pi^2\left(\frac{a}{\delta_T}\right)^2\exp\left(-\frac{4\pi a}{\delta_T}\right)\right).$$

These results do not correspond to (2). Indeed, for (2) we obtain

$$F(a,T)=\frac{\hbar c}{4\pi d^3\delta_T}\left[\varsigma(3)+\sum_{m=1}^{\infty}\sum_{n=1}^{\infty}\exp\left(-\frac{4\pi nma}{\delta_T}\right)S_{mn}(a,\delta_T)\right], \tag{6}$$

$$S_{mn}(a,\delta_T)=\left(\frac{16\pi^2n^2a^2}{m\delta_T^2}+\frac{8\pi na}{m^2\delta_T}+\frac{2}{m^3}\right).$$

At high temperatures $a/\delta_T >> 1$, and it follows from (6) that $F(a,T)=\hbar c(\varsigma(3)+\exp(-4\pi a/\delta_T)S_{11}(a,\delta_T))/4\pi d^3\delta_T$. It differs from (5). At low temperatures $a/\delta_T << 1$, the indices in (6) run infinitely close to values, and the double sum can be calculated by reducing it to integrals. When $a/\delta_T \to 0$ it diverges, and in order to comply with (1), in order to obtain Casimir's law, it must have an infinite limit increasing as $\pi^3\delta_T/(60a)$. Denoting $4\pi mna/\delta_T = z_m$, we see that the points are located infinitely close, and the sum over *m*, using $dm=\delta_T dz/(4\pi na)$, can be replaced by an integral:

$$\sum_{m=1}^{\infty}=(4\pi n)^2\frac{d^2}{\delta_T^2}\int_{\frac{4\pi na}{\delta_T}}^{\infty}\exp(-z)\left(\frac{1}{z}+\frac{2}{z^2}+\frac{2}{z^3}\right)dz\,.$$

Let Integrate the first term by parts. It reduces the second term by a factor of two and gives an extraintegrative contribution. The integral of the parts of the second term $z^{-2}$ cancels out the third term, and the result remains

$$F(a,T)=\frac{\hbar c}{4\pi a^3\delta_T}\left[\varsigma(3)+\sum_{n=1}^{\infty}\exp\left(-\frac{4\pi na}{\delta_T}\right)\left(\frac{4\pi na}{\delta_T}+1\right)\right].$$

In it, the variable *n* also runs through an almost continuous range of values, therefore, using $dn=\delta_T dz/(4\pi a)$ and converting the sum into an integral, we obtain

$$F(a,T)\approx\frac{\hbar c}{4\pi a^3\delta_T}\left(\varsigma(3)+2+\frac{\delta_T}{2\pi a}\exp\left(-\frac{4\pi a}{\delta_T}\right)\right)\ . \qquad (7)$$

This result tends to $\hbar c/\left(8\pi^2 a^4\right)$, which is about three times less than Casimir's result. So, the formula (2) does not match to (1). This is due to the fact that (2) does not take into account the contour integral, i.e. this formula approximately corresponds to the case of extremely high temperatures. Lifshitz in [3] explained the absence of a contour integral in this way: “integration along the sections of the imaginary axis between the poles gives purely imaginary values that fall out when taking Re.” However, when using the argument principle, the function is considered analytical outside the poles, the real part is not taken, and the integral along the contour is equal to the value $2\pi i$ multiplied by the sum of zeros minus the sum of the poles: $(2\pi i)^{-1}\int_G df/f=\sum_\alpha \mathrm{res}_\alpha = N-P$ [6,10]. When the poles hit the contour (as in our case), the contour is changed by introducing infinitesimal semicircles, and the pole points should be considered as gouged out and the Newton-Leibniz formula should be applied, i.e. the integral should be considered in the sense of the Cauchy principal value. Zeros when calculating energy do not affect the result. Since the residuals are imaginary and the integral along the imaginary axis is imaginary, the result is real. In this case, the cotangent poles are on the imaginary axis, and the sum of the half-rings plus the contour integral between the poles should be taken.

Further in (3) it is stated that (2) passes into (1) at $T\to 0$. Let's check it out. We have $\delta_T\to\infty$, the points become closely spaced, and it is indeed possible to move from the sum $\sum(dn)$ to the integral $\hbar c/(2\pi)\int dk$ if $dn=dk\delta_T/(2\pi)\to 0$. However, this result is not valid for small distances, since the condition $\delta_T\to\infty$ is not compatible with extremely small $a$ when the integral diverges. As shown above, the limit depends on the ratio $\delta_T/a$ of the parameters. At low $a$ and at the finite temperature, the equation (2) does not correspond to (1) and may differ significantly. To verify this numerically, calculations were performed in Fig. 2 using formula (6) corresponding to (2). 500 terms were used in each of the sums, which is quite enough. It can be seen that at low $a$ and low temperatures, the results are several times less than the Casimir result. For large $a$, the results go above the Casimir result and tend to the formula $F(a,T)=\hbar c\varsigma(3)/\left(4\pi d^3\delta_T\right)$. It is interesting to note that there is a narrow region (about 1μm) where the curves intersect and also intersect with the Casimir result.

So, when calculating the Casimir energy with a function $\coth(\omega\delta_T/2c)=i\cot(k\delta_T/2)$, it has poles $k_n$ on the imaginary axis, taking half-counts in which leads to (2). By swapping variables $z=k\delta_T/2=\xi\delta_T/(2c)$, we reduce the contour integral to the form

$$\tilde{F}(a) = -\frac{4\hbar c}{\pi^2 \delta_T^4}\, p.v.\int_{-\infty}^{\infty} z^3 \cot(z) \int_1^{\infty} \sum_{\mu=e,h} \frac{r_\mu^2 \exp(-4pza/\delta_T)}{1 - r_\mu^2 \exp(-4pza/\delta_T)} p^2 dp dz\,. \tag{8}$$

The minus in (8) is taken because $\coth(iz) = -i\coth(z)$. The integral (8) for $k_p \to \infty$, can be represented as

$$\tilde{F}(d) = -\frac{\hbar c}{\pi^2}\left(\frac{2}{\delta_T}\right)^4 p.v.\int_0^{\infty} z^3 \cot(z) \sum_{m=1}^{\infty} \int_1^{\infty} p^2 \exp(-a_m zp) dp dz\,, \tag{9}$$

where $a_m = 4ma/\delta_T$. The $z^3 \cot(z)\exp(-a_m zp)$ function has poles at the points $z_l = l\pi$, $l = 1,2,...$ In their neighborhood, it can be approximated as $z_l^3 \exp(-a_m z_l p)/(z - l\pi)$. The integral of this function over the domain $(l\pi - \pi/2, l\pi + \pi/2)$ in the sense of the main value is zero. Cotangent is a periodic function having decomposition

$$\cot(\pi x) = \frac{1}{\pi}\sum_{m=-\infty}^{\infty} \frac{1}{x-n}\,, \tag{10}$$

therefore, the integral can be transformed as a sum of integrals over the domain $(0, \pi/2)$ and the infinite number of domains $(n\pi/2, (n+1)\pi/2)$, n=1,2,…: $z = x + \pi/2 + n\pi$

$$\begin{gathered} \tilde{F}(a) = -\frac{\hbar c}{\pi^2}\left(\frac{2}{\delta_T}\right)^4 p.v.\int_0^{\pi/2} z^3\left(\cot(z) - \frac{1}{z}\right)\sum_{m=1}^{\infty}\int_1^{\infty} p^2 \exp(-a_m pz) dp dz + \\ + \frac{\hbar c}{\pi^2}\left(\frac{2}{\delta_T}\right)^4 p.v.\sum_{n=1}^{\infty}\sum_{m=1}^{\infty}\int_0^{\pi} f_{mn}(z,p)\tan(z) dp dz \end{gathered}\,. \tag{11}$$

Here $b_{mn} = 4\pi m(n - 1/2)a/\delta_T$, $f_{mn}(z,p) = (z + (n-1/2)\pi)^3 p^2 \exp(-(a_m z + b_{mn})p)$. In (11), we subtracted 1/$z$ from the cotangent. This function becomes zero when integrated over the region $(-\pi/2, \pi/2)$ that was implied when transitioning from the integral over the region $(-\infty, \infty)$ to the region $(0, \infty)$. The cotangent in this domain can be approximated fairly accurately by a function $\cot(z) \approx 1/z + 1/(z-\pi)$, neglecting the deleted terms in the sum (10). Since the integral of the function $1/(z - \pi/2)$ in the sense of the principal value is zero over the domain $(0, \pi)$, we use a function with a removed singularity to calculate the principal value

$$g_{mn}(z,p) = f_{mn}(z,p)\tan(z) - f_{mn}(\pi/2, p)/(z - \pi/2)\,. \tag{12}$$

Let us denote there $G_{mn}(z)$ is an integral of the function $f_{mn}(z,p)$ with respect to $p$:

$$\begin{gathered} G_{mn}(z) = \exp(-(a_m z + b_{mn}))(z + (n-1/2)\pi)^3 \times \\ \times\left[\frac{1}{a_m z + b_{mn}} + \frac{2}{(a_m z + b_{mn})^2} + \frac{2}{(a_m z + b_{mn})^3}\right] \end{gathered}\,. \tag{13}$$

Then the integral of (12) is $\tilde{G}_{mn}(z) = G_{mn}(z)\tan(z) - G_{mn}(\pi/2)/(z-\pi/2)$. Let us denote the result of integrating this function with respect to $z$ as $\tilde{\tilde{G}}_{mn}$. We also denote the result of integrating $\exp(-a_m pz)$ with respect to $p$ as

$$\tilde{g}_m(z) = \exp(-(a_m z))\left[\frac{1}{a_m z} + \frac{2}{(a_m z)^2} + \frac{2}{(a_m z)^3}\right], \tag{14}$$

and the result of the integration $z^3(\cot(z) - 1/z)\tilde{g}_m(z)$ is as follows $\tilde{\tilde{g}}_m$. Then the contribution from the contour integral takes the form

$$\tilde{F}(a) = -\frac{\hbar c}{\pi^2}\left(\frac{2}{\delta_T}\right)^4 \sum_{m=1}^{\infty}\left(\tilde{\tilde{g}}_m - \sum_{m=1}^{\infty}\tilde{\tilde{G}}_{mn}\right). \tag{15}$$

It is quite simple to calculate it numerically, since all the peculiarities have been eliminated. An example of such a calculation is shown in Fig. 2, dashed line 4. For small $a$, it lies above the value given by formula (6) and approaches Casimir's result. For extremely large $a$, it practically does not differ from (6). For a qualitative assessment of (15), we can take approximations of the cotangent and tangent and calculate the integral for extremely small and extremely large $a$. For the cotangent in the region $(0, \pi/2)$, we can take $\cot(z) \approx 1/z + 1/(z-\pi)$, and for the tangent in the region $(0, \pi)$ we can take $\tan(z) \approx -2/(2z-\pi) + 2\,\mathrm{sgn}(2z-\pi)/\pi$. Then, to obtain $\tilde{\tilde{g}}_m$, we should integrate the function.

$$\frac{\exp(-a_m z)}{z-\pi}\left[\frac{z^2}{a_m} + \frac{2z}{a_m^2} + \frac{2}{a_m^3}\right].$$

The integral can be estimated by taking the average value $1/(z-\pi) \approx -4/(3\pi)$. Then

$$\tilde{\tilde{g}}_m = -\frac{8}{\pi a_m^4} + \left(\frac{\pi}{3a_m^2} + \frac{8}{3a_m^3} + \frac{8}{\pi a_m^4}\right)\exp(-a_m \pi/2).$$

At short distances, we get

$$\tilde{\tilde{g}}_m \approx -\frac{4}{3a_m^3} + \frac{\pi}{3a_m^2} \approx -\frac{4}{3a_m^3}.$$

At large distances, we have

$$\tilde{\tilde{g}}_m \approx -\frac{8}{\pi a_m^4} + \frac{\pi}{3a_m^2}\exp(-a_m \pi/2),$$

and we obtain a very small contribution to the contour integral from this term. After rather cumbersome calculations, by expanding in the neighborhood of $z \approx \pi/2$, it can be shown that for extremely small $a$ we have $\tilde{\tilde{G}}_{mn} \sim 1/a_m^4$, i.e., the contour integral increases the result (7). In this case, we can take $G_{mn}(z) \approx 2(n\pi)^3 \exp(-(a_m z + b_{mn}))/(a_m z + b_{mn})^3$. Then

$$\tilde{G}_{mn}(z) \approx -\frac{G_{mn}(z) - G_{mn}(\pi/2)}{z - \pi/2} + 2G_{mn}(z)\operatorname{sgn}(2z - \pi)/\pi .$$

The first term can be replaced with $-\partial_z G_{mn}(z)_{z=\pi/2}$, i.e., the integral of it is equal to $\pi\partial_z G_{mn}(\pi/2)$. It has an order of

$$\pi\partial_z G_{mn}(\pi/2) \approx \frac{3}{16\pi} \exp(-4\pi mna/\delta_T) \frac{\delta_T^3}{a^3 m^3 n}. \tag{16}$$

The integral of the second term is zero. When summing (16), the series in $n$ diverges logarithmically for $a \to 0$. At the same time, the sum in m exists. Divergence means that the sum tends to $A(a)(\delta_T^3/a^3)\ln(\delta_T/a)$. The coefficient $A(a)$ is greater than one, and in the limit, the equality $A(a)\ln(\delta_T/a) = C\delta_T/a$ may hold, where $C$ is a constant, i.e., the contribution (9) can increase (7) to the Casimir result. For large $a$, the sum with $\tilde{\tilde{G}}_{mn}$ gives a very small correction to (6). Unfortunately, it is difficult to obtain the values of the constants analytically. However, numerical calculations confirm it.

In conclusion, we note that in [3] formulas (1) and (2) are not strictly derived (for this conclusion, see [11]). Formula (1) is brilliantly guessed. In addition to a fairly large number of inaccuracies in the formulation of the problem (see [11]), its conclusion is replaced by the phrase "*After a series of transformations, we can represent $F_\omega$ in the following form*". However, the above form of $F_\omega$ diverges when integrated in frequency, as the author himself noted in [3]. In addition, the formula (2.4) finally given in [3] also differs. Formula (1) gives the final values for any positive $a$. In earlier works, Lifshits also cited these formulas without inference. In [3], the withdrawal of force was assumed by using the Rytov method. The method is based on the introduction of fluctuation sources proportional to dissipation $\varepsilon''(\omega)$ into Maxwell's equations with the expansion of fields into plane waves and the imposition of boundary conditions. Obviously, this approach can only take into account the force associated with thermal fluctuations. Then all the coefficients of the expansions would have to be proportional to $\varepsilon''(\omega)$, and the resulting force would have to be $\varepsilon''(\omega)$ under the sign of the integral, which is not the case in (1). This apparently prompted a number of authors [5,6,12-14] to conclude (1). In addition to the principle of the argument [5,6], the following were used for the conclusion: correlation relations based on the Lorentz lemma and the excitation of structures by dipoles [12], the variational principle with the introduction of Green functions of electrodynamics [13], methods of quantum field theory in statistical physics [14]. The correlation relations for structures are obtained from the problems of their excitation by dipoles [12,15], therefore they depend on the shape of the bodies, and in the case under consideration they should depend on a, which is not present in the formulation of the problem in [3]. The dependence of correlations on the shape of bodies makes it more difficult to obtain results for bodies of arbitrary shape. For some structures, for example, CNTs at a high temperature and a large distance (compared to the radius), thermal fluctuations make a significant contribution to the force, since boundary conditions at a large distance have little effect. Therefore, it is possible to simplify the problem by not introducing excitation by dipoles, but by introducing only thermal fluctuations in bodies as sources of fields in Maxwell's equations [16,17].

The literature discusses the possibility of taking real parts when calculating energy by summing complex frequencies in dissipative structures (see, for example, [3,7]). Here it should be noted once again that in thermodynamic equilibrium, a structure emits and absorbs equally at any frequency, i.e. its free energy is conserved. Formally, this means that losses can be ignored. At thermodynamic equilibrium, the spectrum does not change, energy is not lost, and the van Kampen-Schram method gives the real free energy of the structure.

Figure 1 shows the results of calculating the force (1) for thick metal plates using the Drude model. For distances $a < \lambda_p = 2\pi c / \omega_p$, the Drude model stops working, so three terms of the Lorentz model (curve 2) with resonant frequencies are added to it so that at low frequencies the value $\varepsilon_L = 9.6$ is obtained. Accordingly, at high frequencies $\varepsilon_L \to 1$. Figure 2 shows the calculation results according to formula (6) corresponding to (2) for $r_\mu = 1$. Note that the discrepancy between (1) and (2) at low temperatures and distances is also shown by numerical calculations using these formulas for real media with specified DP models. We do not present these results because the case of perfectly conductive screens is the simplest and most convincing.

**Financing the work**

The study was carried out with the financial support of the Ministry of Education and Science of the Russian Federation as part of a state assignment (project No. FSRR-2026-0006).

## References

[1 ] H.B.G. Casimir. On the attraction between two perfectly conducting plates. Proc. K. Ned. Akad. Wet. **51**, 793–795 (1948).

[2 ] E.M. Lifshitz. The Theory of Molecular Attractive Forces between Solids. Sov. Phys. JETP **2**, 73–83 (1956).

[3 ] S.K. Lamoreaux. The Casimir force: Background, experiments and applications. Reps. Progr. Phys. **65**, 201–236, (2005). DOI:10.1088/0034-4885/68/1/R04.

[4 ] U. Leonhardt. *Forces of the quantum vacuum: an introduction to Casimir physics* / editor W.M.R. Simpson (World Scientific Publishing Co. Pte. Ltd., Singapore, 2015).

[5 ] N.G. Van Kampen, B.R.A. Nijboer, K. Schram, On the macroscopic theory of van der Waals forces, Phys. Lett. A **26**, 307–308 (1968). DOI: 10.1016/0375-9601(68)90665-8.

[6 ] K. Schram. On the macroscopic theory of retarded Van der Waals forces. Physics Letters A **43**(3), 282–284 (1973). DOI: 10.1016/0375-9601(73)90307-1%20.

[7 ] F. Intravaia, R. Behunin. Casimir effect as a sum over modes in dissipative systems. Phys. Rev. A **86**, 062517 (2012) DOI: 10.1103/PhysRevA.86.062517.

[8 ] F. Intravaia. How modes shape Casimir physics. International Journal of Modern Physics A **37**(19), 2241014 (2022). DOI: 10.1142/S0217751X22410147.

[9 ] J. Mehra. Temperature correction to Casimir effect. Physica **37**, 145–152 (1967). DOI: 10.1016/0031-8914(67)90115-2.

[10 ] B.V. Shabbat. Introduction to complex analysis. Nauka, Moscow (1976). [In Russian].

[11 ] M.V. Davidovich. On the Lifshitz formula of dispersion interaction. arXiv:2008.0025 (2026).

[12 ] M.V. Levin, S.M. Rytov. Theory of equilibrium thermal fluctuations in electrodynamics. Nauka, Moscow (1967). 308 p. [In Russian].

[13 ] J. Schwinger, L.L. DeRaad, K.A. Milton. Casimir effect in dielectrics, Ann. Phys. **115**(1), 676–698 (1978). DOI: 10.1016/0003-4916(78)90172-0.

[14 ] I.E. Dzyaloshinskii, E.M, Lifshitz, L.P. Pitaevskii. The general theory of van der Waals forces. Advances in Physics, **10**(38). 165–209 (1961). DOI: 10.1080/00018736100101281.

[15 ] M.V. Davidovich. Correlations, thermal radiation and absorption for graphene. Technical Physics **96**(7), 1316–1327 (2026). DOI: 1061011/JTP.2026.07.63120.298-25.

[16 ] I.S. Nefedov, M.V. Davidovich, O.E. Glukhova, M.M. Slepchenkov, J.M. Rubi. Radiative heat transfer between two carbon nanotubes. Sci. Rep. **12**(I), 17930 (2022). https://doi.org/10.1038/s41598-022-22138-8.

[17 ] I.S. Nefedov, M.V. Davidovich, O.E. Glukhova, M.M. Slepchenkov, M. Rubi. Casimir forces between two carbon nanotubes. Phys. Rev. B **104**(2), 085409-8 (2021). DOI: 10.1103/PhysRevB.104.085409.

Figures

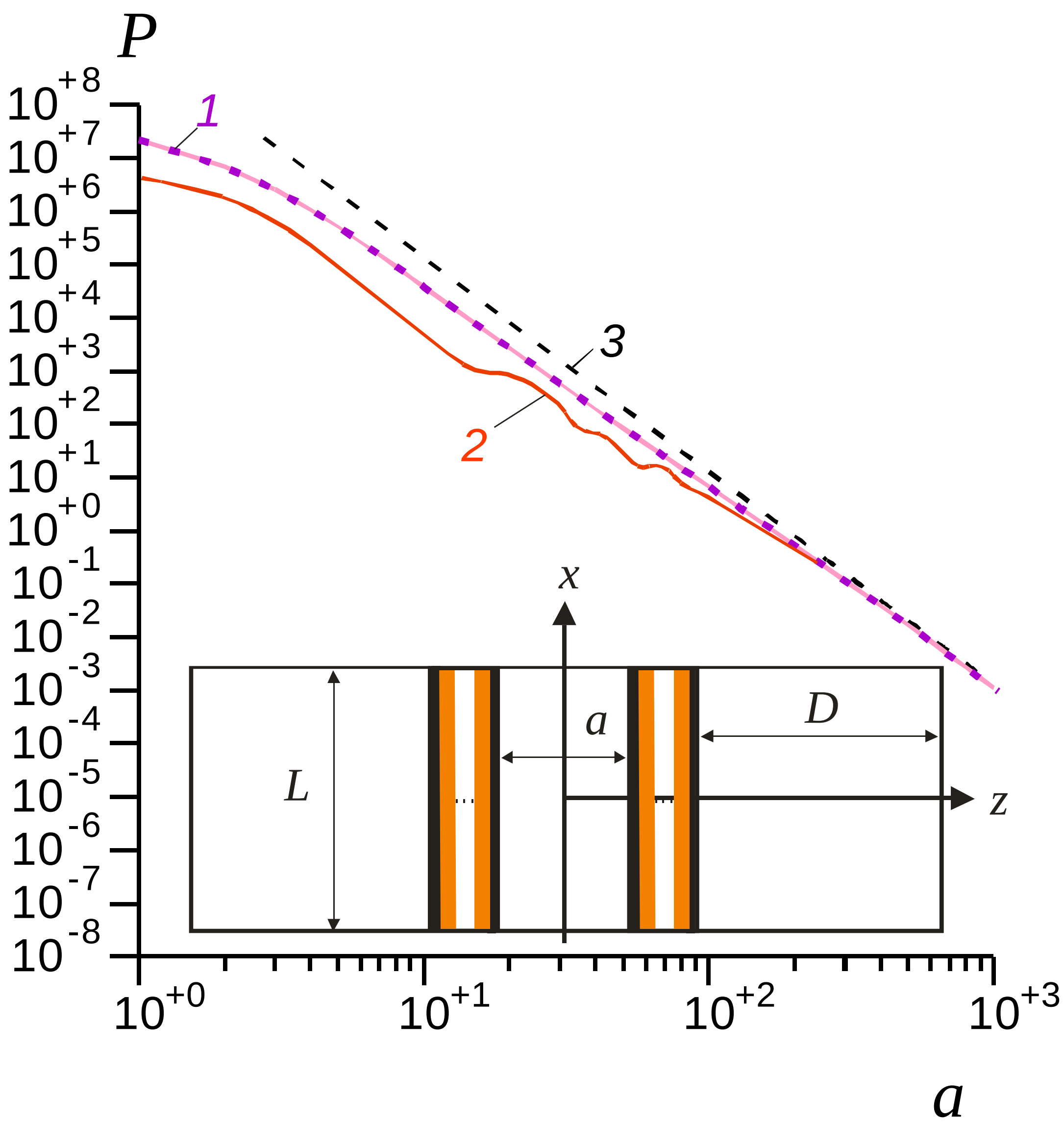

Fig. 1. Configuration of the structure (insert) and the external Casimir pressure $P$ (N/m$^2$) depending on the size $a$ (nm) for thick metal plates (Drude model, silver, $\omega_p$=1.57×10$^{16}$ Hz, $\omega_c$=3.2×10$^{13}$ Hz, $\varepsilon_L$=9.6), separated by a vacuum gap $a$ (curve 1) and approximated by $\varepsilon_L$ by three Lorentz terms with frequencies of 2×10$^{17}$, 5×10$^{17}$ and 10$^{18}$ Hz (2). Casimir's result is shown by dashed line 3

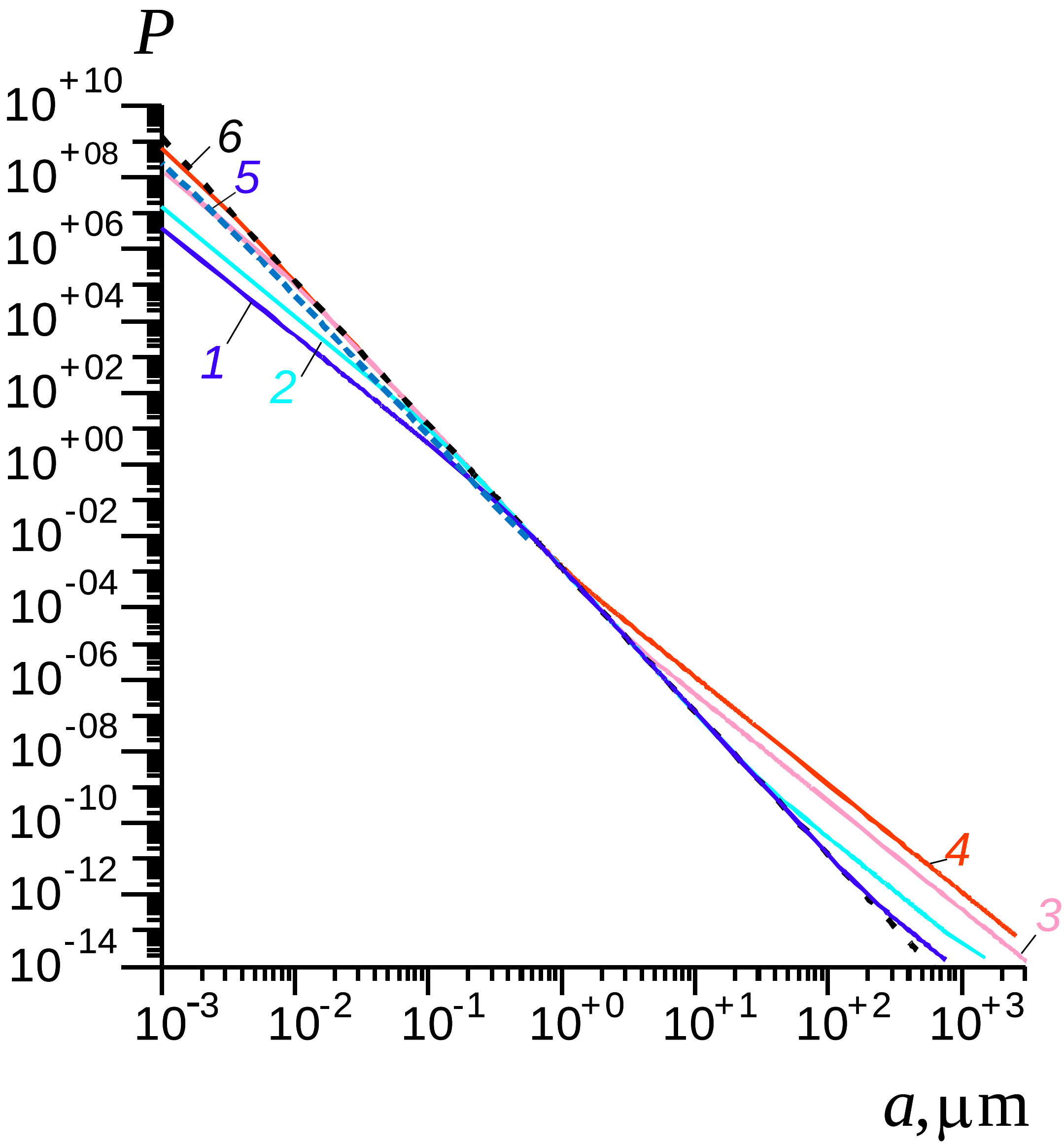

Fig. 2. External Casimir pressure $P$ (N/m$^2$) on two ideally conductive metal layers depending on $a$ (in μm) at different temperatures (formula (6)): 5 K (curve 1), 30 K (2), 300 K (3) and 900 K (4). The curve 5 is the correction (6) using (15). Casimir's result is a dashed line 6